\documentclass[preprint,aps,prc,tightenlines,nofootinbib,showpacs,amssymb]{revtex4}
\begin{document}
\input{psfig}
\input{epsf}
\def\Im{\mbox{\sl Im\ }}
\def\pd{\partial}
\def\oln{\overline}
\def\olft{\overleftarrow}
\def\ds{\displaystyle}
\def\bgreek#1{\mbox{\boldmath $#1$ \unboldmath}}
\def\sla#1{\slash \hspace{-2.5mm} #1}
\newcommand{\bra}{\langle}
\newcommand{\ket}{\rangle}
\newcommand{\vep}{\varepsilon}
\newcommand{\met}{{\mbox{\scriptsize met}}}
\newcommand{\lab}{{\mbox{\scriptsize lab}}}
\newcommand{\cm}{{\mbox{\scriptsize cm}}}
\newcommand{\mcal}{\mathcal}
\newcommand{\Del}{$\Delta$}
\newcommand{\g}{{\rm g}}
\long\def\Omit#1{}
\long\def\omit#1{\small #1}
\def\beq{\begin{equation}}
\def\eeq{\end{equation} }
\def\bea{\begin{eqnarray}}
\def\eea{\end{eqnarray}}
\def\eqref#1{Eq.~(\ref{eq:#1})}
\def\eqlab#1{\label{eq:#1}}
\def\figref#1{Fig.~\ref{fig:#1}}
\def\figlab#1{\label{fig:#1}}
\def\tabref#1{Table \ref{tab:#1}}
\def\tablab#1{\label{tab:#1}}
\def\secref#1{Section~\ref{sec:#1}}
\def\seclab#1{\label{sec:#1}}
\def\VYP#1#2#3{{\bf #1}, #3 (#2)}  
\def\NP#1#2#3{Nucl.~Phys.~\VYP{#1}{#2}{#3}}
\def\NPA#1#2#3{Nucl.~Phys.~A~\VYP{#1}{#2}{#3}}
\def\NPB#1#2#3{Nucl.~Phys.~B~\VYP{#1}{#2}{#3}}
\def\PL#1#2#3{Phys.~Lett.~\VYP{#1}{#2}{#3}}
\def\PLB#1#2#3{Phys.~Lett.~B~\VYP{#1}{#2}{#3}}
\def\PR#1#2#3{Phys.~Rev.~\VYP{#1}{#2}{#3}}
\def\PRC#1#2#3{Phys.~Rev.~C~\VYP{#1}{#2}{#3}}
\def\PRD#1#2#3{Phys.~Rev.~D~\VYP{#1}{#2}{#3}}
\def\PRL#1#2#3{Phys.~Rev.~Lett.~\VYP{#1}{#2}{#3}}
\def\FBS#1#2#3{Few-Body~Sys.~\VYP{#1}{#2}{#3}}
\def\AP#1#2#3{Ann.~of Phys.~\VYP{#1}{#2}{#3}}
\def\ZP#1#2#3{Z.\ Phys.\  \VYP{#1}{#2}{#3}}
\def\ZPA#1#2#3{Z.\ Phys.\ A\VYP{#1}{#2}{#3}}
\def\half{\mbox{\small{$\frac{1}{2}$}}}
\def\quarter{\mbox{\small{$\frac{1}{4}$}}}
\def\nn{\nonumber}
\newlength{\PicSize}
\newlength{\FormulaWidth}
\newlength{\DiagramWidth}
\newcommand{\vslash}[1]{#1 \hspace{-0.42 em} /}
\newcommand{\qslash}[1]{#1 \hspace{-0.46 em} /}
\def\her{\marginpar{$\Longleftarrow$}}
\def\bel{\marginpar{$\Downarrow$}}
\def\abo{\marginpar{$\Uparrow$}}



\title{Contribution of spin $1/2$ and $3/2$ resonances to
two-photon exchange effects in elastic electron-proton scattering}

\author{S.~Kondratyuk}
\affiliation{Department of Physics and Astronomy, University of Manitoba,
Winnipeg, MB, Canada R3T 2N2}
\author{P.~G.~Blunden}
\affiliation{Department of Physics and Astronomy, University of Manitoba,
Winnipeg, MB, Canada R3T 2N2}

\date{\today}

\begin{abstract}

We calculate contributions of hadron resonances to
two-photon exchange effects in electron-proton scattering.
In addition to the nucleon and $P33$ resonance, the following heavier resonances 
are included as intermediate states in the two-photon exchange diagrams:
$D13$, $D33$, $P11$, $S11$ and $S31$. 
We show that the corrections due to the heavier resonances 
are smaller that the dominant nucleon and $P33$ contributions.
We also find that there is a partial cancellation between the contributions from
the spin $1/2$ and spin $3/2$ resonances, which results in a further suppression of their
aggregate two-photon exchange effect.

\end{abstract}

\maketitle



In the past few years there has been a remarkable renewal of interest in the nucleon electromagnetic form factors. Its principal motivation has been an effort to interpret the results of recent
polarization transfer electron-proton scattering experiments~\cite{Jon00}: the 
form factors from these measurements 
are in strong disagreement with those obtained from the unpolarized
scattering (Rosenbluth cross section) data~\cite{Wal94,Qat05}, if
one uses the standard one-photon exchange approximation to extract the form factors.
This problem has to be resolved -- not only in view of the fundamental importance of the
nucleon form factors, but also 
in order that electron-proton scattering should be 
used as a reliable tool to measure them.
A comprehensive up-to-date review of the subject can be found in~\cite{Arr06} and in
references cited therein.

It was shown in 
Refs.~\cite{Blu03,Gui03,Che04} that by taking into account two-photon exchange diagrams one can reconcile 
the Rosenbluth and polarization transfer measurements of the proton electromagnetic form factors. 
Since the $\Delta$ resonance plays an important role in
nucleon Compton scattering, its contribution was investigated in~\cite{Kon05}, where it was 
demonstrated that it is essential that both the nucleon and the $\Delta$ intermediate states 
be included in evaluating two-photon exchange effects in electron-proton scattering. 

The present report extends the approach of Refs.~\cite{Blu03,Kon05}, generalizing it to 
include the full spectrum of the most important hadron resonances as intermediate states
in the two-photon exchange box- and crossed-box loop diagrams for electron-proton scattering.
We will use computational techniques quite similar to those
described in the earlier papers~\cite{Blu03,Kon05}. Therefore, in this report we will
give only the details which are specifically relevant to the extension to the
general spin $1/2$ and $3/2$ resonances. 
We take the masses of the resonances and their nucleon-photon coupling constants 
based on dynamical multichannel calculations~\cite{Kon01,Kor00} of nucleon
Compton scattering at low and intermediate energies. The resonance two-photon exchange effects turn out to be not too sensitive to the details of these models.

We will show that in general the contributions of all the heavier resonances are much smaller than those of the nucleon and $\Delta$ ($P33$) calculated earlier~\cite{Kon05}. We will analyse the contributions of the individual resonance in some detail. In particular,
the calculations presented below will reveal an interesting interplay between the
contributions of the spin $1/2$ and spin $3/2$ resonances, which is analogous to the partial
cancellation of the two-photon exchange effects of the nucleon and $\Delta$
intermediate states, found in Ref.~\cite{Kon05}. 
One of the results of this report is that,
notwithstanding the smallness of the resonance contributions, their inclusion
in the two-photon exchange diagrams leads to a better agreement between the
Rosenbluth and polarization transfer data analyses, especially at higher values of the
momentum-transfer squared $Q^2$.


The differential cross section for elastic electron-proton scattering can be written as
\beq
\sigma=\sigma_B (1+\delta),
\eqlab{cs}
\eeq
where $\delta$ stands for a two-photon exchange correction to the Born one-photon exchange cross section $\sigma_B$. Throughout this paper
we will consider the reduced cross section, defined in the 
standard way~\cite{Arr06,Kon05} by
omitting an irrelevant (for the present purposes) 
factor describing the scattering on a structureless spin $1/2$ target.
The Born contribution to the reduced cross section is given in terms of the electric and magnetic form factors of the proton,
$G_E(Q^2)$ and $G_M(Q^2)$, as follows:
\beq
\sigma_B = G_M^2(Q^2) + {\epsilon \over \tau} G_E^2(Q^2)\,.
\eqlab{xsec_red}
\eeq
The two independent kinematical variables are 
the momentum transfer squared
$Q^2 \equiv -q^2 \equiv 4 \tau M^2$ ($M$ is the nucleon mass)
and the photon polarization $ \epsilon= [ 1+2( 1+\tau ) \tan^2(\theta/2) ]^{-1} $, 
the latter expressed in terms of the scattering angle $ \theta $.
The various contributions to the $\delta$ in~\eqref{cs} can be calculated 
from the scattering amplitudes ${\mathcal{M}}$ using~\cite{Blu03}
\beq
\delta_{N, R} =
2 \frac{\mbox{Re} \left({\mathcal{M}}_B^\dagger \, 
{\mathcal{M}}_{N, R}^{\gamma \gamma} \right)}{ \left| {\mathcal{M}}_B \right|^2 }\,,
\eqlab{del_nr}
\eeq
where the subscript $B$ (superscript $\gamma \gamma$) denotes the
Born (two-photon exchange) contribution. The two-photon exchange 
box and crossed box-diagrams can have
nucleon and resonance intermediate states, denoted in~\eqref{del_nr} by the
subscripts $N$ and $R$. To leading order in the electromagnetic coupling, the total two-photon
exchange correction is given by the sum of the separate hadron contributions:
\beq
\delta=\delta_N+\delta_{P33}+\delta_{D13}+\delta_{D33}+\delta_{P11}+
\delta_{S11}+\delta_{S31} \,.
\eqlab{del}
\eeq

The coupling of a spin $3/2$ resonance (mass $M_R$) to a nucleon and a photon is described by the vertex~\footnote{We use the conventions of Ref.~\cite{Bjo64}.}
\bea
&\Gamma_{\gamma R \rightarrow N}^{\nu \alpha}(p,q) 
= i {\ds \frac{e F_R(q^2)}{2 M_R^2}} \bigg\{
g_1^R \left[\, g^{\nu \alpha} \vslash{p} \qslash{q} - p^\nu \gamma^\alpha \qslash{q} 
- \gamma^\nu \gamma^\alpha p \cdot q + \gamma^\nu \vslash{p} q^\alpha\, \right]  & \nn \\
& + g_2^R \left[\, p^\nu q^\alpha - g^{\nu \alpha} p \cdot q\, \right] 
+ (g_3^R/M_R) \left[\,q^2 (p^\nu \gamma^\alpha - g^{\nu \alpha} \vslash{p}) +
q^\nu (q^\alpha \vslash{p} - \gamma^\alpha p \cdot q )\, \right] 
\bigg\} P_R\, I_R\,, 
\eqlab{vert32}
\eea
where $p^\alpha$ and $q^\nu$ are the four-momenta of the resonance and photon, respectively,
and $g_{1,2,3}^R$ are coupling constants discussed below.
The Lorentz factor $P_R=\gamma_5$ if $R=P33$, and $P_R=1$ if $R=D13$ or $R=D33$;  
and the isospin factor $I_R=T_3$ if $R=P33$ or $R=D33$, and $I_R=1$ if $R=D13$. 

The vertices of the spin $1/2$ resonances read
\beq
\Gamma_{\gamma R \rightarrow N}^{\mu}(q) 
 = - {\ds \frac{e g^R F_R(q^2)}{2 M}} \sigma^{\mu \nu} q_\nu P_R\, I_R\,,
\eqlab{vert12}
\eeq
where for $R=P11$: $P_R=1$, $I_R=1$; for $R=S11$: $P_R=\gamma_5$, $I_R=1$; and for
$R=S31$: $P_R=\gamma_5$, $I_R=T_3$.

The phenomenological form factors $F_R(q^2)$ account for the nonlocal nature of the hadrons while ensuring ultraviolet convergence of the loop integrals. We take the dipole form factors
\beq
F_R(q^2) = \frac{\Lambda_R^{4}}{\left(\Lambda_R^2 - q^2\right)^2}\,,
\eqlab{ff}
\eeq
where $\Lambda_R$ is the cutoff. To keep the number of the
parameters to the minimum, we choose $\Lambda_R = 0.84$ GeV
for all hadrons in the model. This value is known to be consistent with the mean square radius of the proton. 
Taking the same $\Lambda_R$ for all hadrons can be justified since the dependence of the 
two-photon correction
on the form factors is partially cancelled in the ratio~\eqref{del_nr}. 

The vertices in~\eqref{vert32} are orthogonal not only to the photon 4-momentum $q_\nu$ 
(the usual gauge invariance property), but also to the resonance 4-momentum $p_\alpha$.
The latter property ensures the possibility of using only the physical spin $3/2$
part of the Rarita-Schwinger propagator for these resonances,
\beq
S_{\alpha \beta}(p) = \frac{-i}{\vslash{p}-M_R+i0} {\mathcal{P}}^{3/2}_{\alpha \beta}(p)\,,
\eqlab{prop32}
\eeq
with the spin $3/2$ projection operator 
\beq
{\mathcal{P}}^{3/2}_{\alpha \beta}(p) = g_{\alpha \beta} - {1 \over 3} \gamma_\alpha
\gamma_\beta- {1 \over 3 p^2} \left( \vslash{p} \gamma_\alpha p_\beta + 
p_\alpha\gamma_\beta \vslash{p} \right)\,,
\eqlab{proj}
\eeq
while the background spin $1/2$ part of the general propagator does not contribute to the amplitudes~\cite{Pas99}. 

For the spin $1/2$ resonances we use the usual Dirac propagators 
\beq
S(p) = \frac{i}{\vslash{p}-M_R +i0} \,.
\eqlab{prop12}
\eeq
At present, we neglect the widths of the resonances. While entering into the imaginary part of the two-photon exchange amplitude, these widths do not directly affect the real part and
hence, by~\eqref{del_nr}, they should not be significant in the calculation of $\delta_R$.

We use the following masses of the resonances
(in units of GeV)~\cite{Kor00}:
$M_{P33}=1.232$, $M_{D13}=1.52$, $M_{D33}=1.7$, $M_{P11}=1.55$,
$M_{S11}=1.535$, $M_{S31}=1.62$, and $M=0.938$ for the nucleon.
As was done previously for the $\Delta$ resonance~\cite{Kon05}, 
we choose the coupling constants in the vertices Eqs.~(\ref{eq:vert32}) and (\ref{eq:vert12})
using the Dressed K-Matrix Model (DKM) whose essential ingredients are described 
in~\cite{Kon01}.
The results discussed below were obtained using the following coupling constants:
$g_1^{P33}=7$, $g_2^{P33}=9$,
$g_1^{D13}=g_2^{D13}=g_1^{D33}=g_2^{D33}=0.1$, $g^{P11}=1.2$,   
$g^{S11}=-0.45$, $g^{S31}=-0.2$. With these numerical values for the constants, the DKM provides a good  description of nucleon Compton scattering at energies from zero up to the second resonance region. Note however that a precise tuning of the resonance coupling constants is unnecessary for the purposes of the present calculation; this point will be
explained in more detail below. 
The chosen set of coupling constants implies that the $R \rightarrow \gamma N$ 
transitions are mostly of the magnetic type, with the electric type being much smaller. As was shown in Ref.~\cite{Kon05} for the case of the $\Delta$ intermediate state, 
the two-photon exchange contribution of the Coulomb coupling $g_3^{P33}$ is 
about 10 times smaller than that of the magnetic coupling. 
A similarly strong suppression is anticipated for the other resonances as well. 
Since our main focus is on the dominant two-photon exchange effects,
in this report we omit the Coulomb couplings for all of the resonances.


Using the above vertices and propagators we evaluate the box- and crossed-box two-photon exchange loop diagrams. The calculation is fully relativistic and obeys the properties of
gauge invariance and crossing symmetry. The loop integrals are finite: the infrared convergence is due to the masses of the intermediate resonances being greater than the nucleon mass, and the ultraviolet convergence is ensured by the presence of the regularizing form factor $F_R(q^2)$ in the vertices Eqs.~(\ref{eq:vert32}) and (\ref{eq:vert12}).
The loop integrals and their evaluation involve obvious 
generalizations of the expressions given in Ref.~\cite{Kon05} where all technical details 
can be found.
The sum of the box- and crossed-box loop integrals for each resonance $R$ constitutes the 
two-photon exchange
amplitude ${\mathcal{M}}_{R}^{\gamma \gamma}$. 
The two-photon exchange correction to the unpolarized electron-proton scattering cross section
is then given by~\eqref{del_nr} for each resonance $R$ separately, 
and by~\eqref{del} for the contribution of all hadron intermediate states.


The calculated two-photon corrections to the reduced cross 
section are displayed in~\figref{csred}. 
The one-photon exchange Born cross sections are shown by the dotted lines.
The cross sections including additional two-photon exchange corrections are
shown by the dashed lines for the sum of the nucleon and $P33$ contributions, and by 
the solid lines for the full result with all resonances.
In general, each resonance two-photon correction is  
proportional to a sum of squares of the nucleon-photon coupling constants of that resonance.
This sets the scale of the magnitude of the resonance contributions. 
Taking an example of $Q^2=4$ GeV$^2$,
the two-photon exchange corrections from the included hadrons can be classified by their signs and orders of magnitude as follows. 
As $0 < \epsilon < 1$, the corrections change smoothly between the values: 
\beq
-4.7 \lesssim \delta_N \lesssim 0 \%, \;\;  
1.9 \gtrsim \delta_{P33} \gtrsim 0 \%, \;\;
-0.7 \lesssim \delta_{D13} \lesssim 0 \%,
\eqlab{res_npd}
\eeq
\beq
-0.3 \lesssim \delta_{D33} \lesssim 0 \%, \;\;
-0.15 \lesssim \delta_{P11} \lesssim 0 \%, \;\; 
0.06 \gtrsim \delta_{S11} \gtrsim 0 \%, \;\;
0.01 \gtrsim \delta_{S31} \gtrsim 0 \%,
\eqlab{res_other}
\eeq
listed in the order of decreasing magnitude.

\begin{figure}[!htb]
\centerline{{\epsfxsize 13.7cm \epsffile[15 65 570 430]{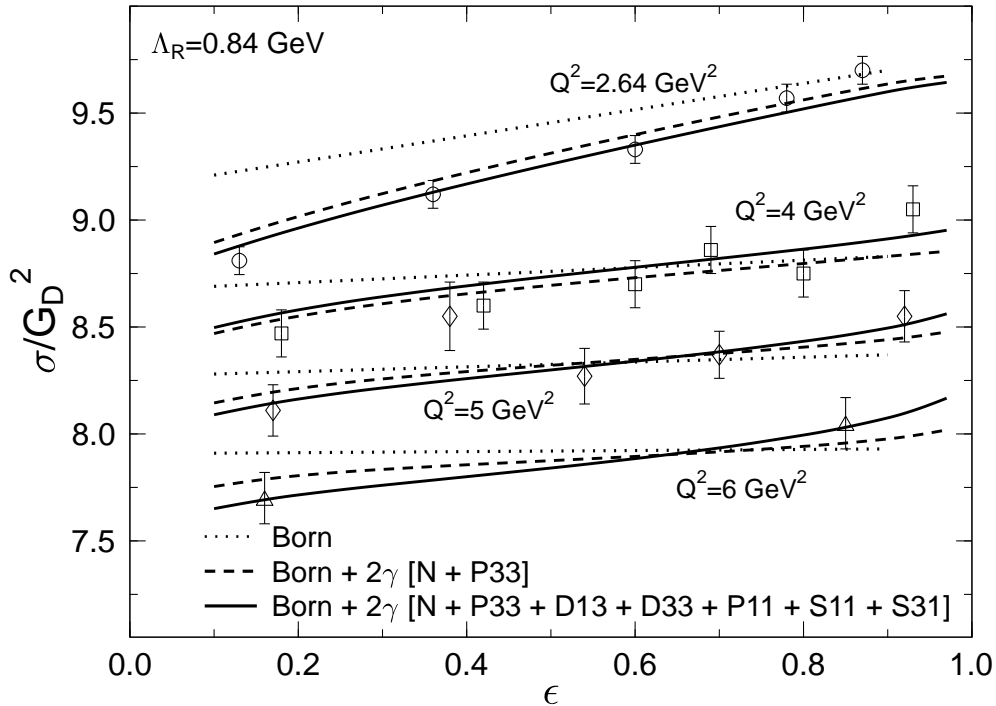}}}
\caption[f3]{
Effect of adding the two-photon exchange correction 
to the Born cross section, the latter evaluated with the
nucleon form factors from the polarization transfer experiment~\cite{Jon00}.
The intermediate state includes a nucleon and indicated hadron resonances.
We show the reduced cross section divided by the square of the standard dipole form factor 
$G_D^2(Q^2)= 1/(1+Q^2/(0.84\,\mbox{GeV})^2)^4$.
The data points at four fixed momentum transfers are taken from Refs.~\cite{Wal94,Qat05}.
\figlab{csred}}
\end{figure}

\figref{csred} shows that at not too high $Q^2$ the two-photon exchange corrections are determined mainly
by the nucleon and $P33$ intermediate states. Therefore, one does not have to fine-tune
the coupling constants of the other resonances to get a good estimate of
the overall two-photon exchange effect. In practice this means that results quite close to those
presented in~\figref{csred} were obtained when we varied the resonance coupling constants 
(except $P33$) by as much as $ \pm 50 \%$.

In addition to the dominant nucleon and $P33$ contributions, 
the $D13$ gives the
most important correction among the remaining resonances. 
This is consistent with the
well-known prominence of the $D13$ in the second resonance region of the Compton scattering cross section, see, e.~g.,~\cite{Kor00} and references therein.
In fact, the fit of the full curves to the data in~\figref{csred} would not be
noticeably worsened if we kept only the nucleon, the $P33$ and the $D13$ 
resonances in the model.  
Even though the nucleon and the $P33$ resonance dominate the two-photon effect, 
the results shown in~\figref{csred} indicate that the inclusion of the
heavier resonances improves the agreement with the data.
This improvement is a genuine dynamical effect since we did not fit 
the calculated cross sections
to the data in~\figref{csred}: rather, the resonance coupling constants were obtained from 
the description of Compton scattering, as described above. 
It it also interesting to note that there is an additional suppression of the aggregate 
two-photon exchange effect of the heavier resonances. Its source is a
partial cancellation between the spin $1/2$ and spin $3/2$ resonance contributions.


It is worth pointing out that the calculation presented in this report is complementary to two sets of existing approaches to two-photon exchange effects in elastic electron-proton scattering.
The first approach~\cite{Blu03,Kon05} takes into account the
intermediate states of the nucleon and the lightest $\Delta$ resonance, which are essential
at all kinematical regimes, but most important at low energies. 
The second approach is based on the generalized parton distribution techniques~\cite{Che04}, 
whose natural application is at relatively high energies.
The energy range between these two sets of models is governed by the
dynamics of the hadron resonances, which has been addressed in the present calculation. 
We have shown that the two-photon exchange effects with the inclusion of the hadron resonances 
are indeed capable of bringing the Rosenbluth and polarization transfer experiments into 
closer agreement with each other. 




\begin{thebibliography}{99}

\bibitem{Jon00} M.~K.~Jones et al., \PRL{84}{2000}{1398}; 
                O.~Gayou et al., \PRL{88}{2002}{092301};
                V.~Punjabi et al., \PRC{71}{2005}{055202} 
                [err.~ibid. \PRC{71}{2005}{069902}.      
\bibitem{Wal94} R.~C.~Walker et al., \PRD{49}{1994}{5671}; 
                L.~Andivahis et al., \PRD{50}{1994}{5491}.
\bibitem{Qat05} I.~A.~Qattan et al., \PRL{94}{2005}{142301}.
\bibitem{Arr06} J.~Arrington, C.~D.~Roberts, and J.~M.~Zanotti, nucl-th/0611050.
\bibitem{Blu03} P.~G.~Blunden, W.~Melnitchouk, and J.~A.~Tjon, 
                \PRL{91}{2003}{142304}.                   
\bibitem{Gui03} P.~A.~M.~Guichon and M.~Vanderhaeghen, \PRL{91}{2003}{142303}.
\bibitem{Che04} Y.~C.~Chen, A.~Afanasev, S.~J.~Brodsky, C.~E.~Carlson, 
                and M.~Vanderhaeghen, \PRL{93}{2004}{122301}.
\bibitem{Kon05} S.~Kondratyuk, P.~G.~Blunden, W.~Melnitchouk, and J.~A.~Tjon, 
                \PRL{95}{2005}{172503}.
\bibitem{Kon01} S.~Kondratyuk and O.~Scholten, \PRC{64}{2001}{024005}.
\bibitem{Kor00} A.~Yu.~Korchin and O.~Scholten, \PRC{60}{2000}{015205};
                G.~Penner and U.~Mosel, \PRC{66}{2002}{055212}.
\bibitem{Bjo64} J.~D.~Bjorken and S.~D.~Drell,
                Relativistic Quantum Mechanics (McGraw-Hill, 1964).
\bibitem{Pas99} V.~Pascalutsa and R.~G.~E.~Timmermans, \PRC{60}{1999}{042201}.

\end{thebibliography}
\end{document}